\documentstyle[12pt]{article}
\def\be{\begin{equation}}
\def\ee{\end{equation}}
\def\l{\label}\def\S{{\cal S}}\def\T{{\cal T}}\def\V{{\cal V}}
\def\H{{\cal H}}\def\K{{\cal K}}\def\Q{{\cal Q}}\def\U{{\cal U}}
\def\V{{\cal V}}\def\W{{\cal W}}\def\s{{\tt s}}\def\t{{\tt t}}

\makeatletter \ifcase\@ptsize \font\teneufm=eufm10
\font\seveneufm=eufm7 \font\fiveeufm=eufm5
\font\teneusm=eusm10 \font\seveneusm=eusm7
\font\fiveeusm=eusm5 \or \font\teneufm=eufm10 scaled
\magstephalf \font\seveneufm=eufm7 \font\fiveeufm=eufm5
\font\teneusm=eusm10 scaled \magstephalf
\font\seveneusm=eusm7 \font\fiveeusm=eusm5 \or
\font\teneufm=eufm10 scaled \magstep1 \font\seveneufm=eufm7
\font\fiveeufm=eufm5 \font\teneusm=eusm10 scaled \magstep1
\font\seveneusm=eusm7 \font\fiveeusm=eusm5 \fi

\newfam\eufmfam \newfam\eusmfam \textfont\eufmfam=\teneufm
\scriptfont\eufmfam=\seveneufm
\scriptscriptfont\eufmfam=\fiveeufm
\textfont\eusmfam=\teneusm \scriptfont\eusmfam=\seveneusm
\scriptscriptfont\eusmfam=\fiveeusm

\def\frak{\ifmmode\let\next\frak@\else
 \def\next{\errmessage{Use \string\frak\space only in math
 mode}}\fi\next} \def\frak@#1{{\frak@@{#1}}}
 \def\frak@@#1{\fam\eufmfam#1} 
 \def\sh{\ifmmode\let\next\sh@\else
 \def\next{\errmessage{Use \string\sh\space only in math
 mode}}\fi\next} \def\sh@#1{{\sh@@{#1}}}
 \def\sh@@#1{\fam\eusmfam#1}

\ifcase\@ptsize \font\tenmsa=msam10 \font\sevenmsa=msam7
 \font\fivemsa=msam5 \font\tenmsb=msbm10
 \font\sevenmsb=msbm7 \font\fivemsb=msbm5 \or
 \font\tenmsa=msam10 scaled \magstephalf
 \font\sevenmsa=msam7 \font\fivemsa=msam5
 \font\tenmsb=msbm10 scaled \magstephalf
 \font\sevenmsb=msbm7 \font\fivemsb=msbm5 \or
 \font\tenmsa=msam10 scaled \magstep1 \font\sevenmsa=msam7
 \font\fivemsa=msam5 \font\tenmsb=msbm10 scaled \magstep1
 \font\sevenmsb=msbm7 \font\fivemsb=msbm5 \fi

\newfam\msafam \newfam\msbfam \textfont\msafam=\tenmsa
\scriptfont\msafam=\sevenmsa
\scriptscriptfont\msafam=\fivemsa \textfont\msbfam=\tenmsb
\scriptfont\msbfam=\sevenmsb
\scriptscriptfont\msbfam=\fivemsb

\def\Bbb{\ifmmode\let\next\Bbb@\else
 \def\next{\errmessage{Use \string\Bbb\space only in math
 mode}}\fi\next} \def\Bbb@#1{{\Bbb@@{#1}}}
 \def\Bbb@@#1{\fam\msbfam#1} \def\hexnumber@#1{\ifnum#1<10
 \number#1\else \ifnum#1=10 A\else\ifnum#1=11
 B\else\ifnum#1=12 C\else \ifnum#1=13 D\else\ifnum#1=14
 E\else\ifnum#1=15 F\fi\fi\fi\fi\fi\fi\fi}
 \def\msa@{\hexnumber@\msafam} \def\msb@{\hexnumber@\msbfam}
 \mathchardef\square="0\msa@03

\makeatother
 \newcommand{\RR}{{\Bbb R}}
\newcommand{\CC}{{\Bbb C}}

\begin{document}\begin{titlepage}

\rightline{UFIFT-HEP-96-28, DPFD96/TH/59}
\rightline{\tt hep-th/9705108}
\rightline{15 May 1997}

\begin{center}

{\Large \bf Quantum Mechanics from an Equivalence Principle}

\vspace{1.cm}

 {\large Alon E. Faraggi$^{1}$ $\,$and$\,$ Marco Matone$^{2}$\\}
\vspace{.2in}
 {\it $^{1}$ Institute for Fundamental Theory, Department of Physics, \\
        University of Florida, Gainesville, FL 32611, USA\\
e-mail: faraggi@phys.ufl.edu\\}
\vspace{.025in}
{\it $^{2}$ Department of Physics ``G. Galilei'' -- Istituto
                Nazionale di Fisica Nucleare\\
        University of Padova, Via Marzolo, 8 -- 35131 Padova, Italy\\
   e-mail: matone@padova.infn.it\\}

\end{center}

\vspace{0.8cm}

\begin{abstract}
We postulate that physical states are equivalent under coordinate
transformations. We then implement this equivalence principle first in the case
of one--dimensional stationary systems showing that it leads to the quantum
analogue of the Hamilton--Jacobi equation which in turn implies the
Schr\"odinger equation. In this context the Planck constant plays the role of
covariantizing parameter. The construction is deeply related to the
$GL(2,{\CC})$--symmetry of the second--order differential equation associated
to the Legendre transformation which selects, in the case of the quantum
analogue of the Hamiltonian characteristic function, self--dual states which
guarantee its existence for any physical system. The universal nature of the
self--dual states implies the Schr\"odinger equation in any dimension.
\end{abstract}

\vspace{.633cm}



\end{titlepage}

\newpage

While  general relativity is based on a simple fundamental principle, a similar
geometrical structure does not seem to underlie quantum mechanics. This
suggests that the problems arising in quantizing gravity are deeply connected
with the apparently different origin of the two theories. In this letter we
postulate that physical systems are equivalent under coordinate transformations.
We will see that while the equivalence principle cannot be consistently
implemented in the Classical Stationary Hamilton--Jacobi Equation (CSHJE), it
leads to its quantum analogue and then to the Schr\"odinger equation. This
quantum stationary Hamilton--Jacobi equation is a third--order differential
equation whose solution defines $\S_0$, denoting the quantum analogue of the
Hamilton characteristic function, or reduced action, $\S_0^{cl}$. Here we
consider the case of stationary one--dimensional systems. The higher
dimensional, time dependent systems will be considered in forthcoming papers.

Our formulation is strictly related to the $GL(2,{\CC})$--symmetry
underlying the recently observed relationship between second--order
differential equations and Legendre transformation. In particular, as we
will see, this identifies in the case of the reduced action $\S_0$, a basic
self--dual state which guarantees the existence of its Legendre transformation
for any system. This is the starting point in a chain of deductions
culminating with the Schr\"odinger equation. In particular, the existence of the
self--dual state implies that for any one--dimensional stationary state with
potential $V$ and energy $E$, there is always a coordinate choice $\tilde q$ for
which $\W(q)\equiv V(q)-E$ corresponds to $\tilde \W (\tilde q)=0$. In this
context the Planck constant, which determines the universal self--dual state,
naturally arises as a covariantizing parameter.

Let us denote by $q$ the coordinate and by $p$ the momentum of a stationary
physical system. We assume the existence of $\T_0$, the Legendre dual of $\S_0$,
that is
\be
p=\partial_q\S_0(q),\qquad q=\partial_p\T_0(p),
\l{pq}\ee
\be
\S_0=p\partial_p\T_0-\T_0.
\l{lt}\ee

Let us consider the $GL(2,{\CC})$--transformations
\be
\tilde q ={Aq+B\over Cq+D},\qquad\tilde p = \rho^{-1}(Cq+D)^2p,
\l{4}\ee
where $\rho=AD-BC\ne 0$. These transformations are equivalent to say that $\S_0$
is $GL(2,{\CC})$--invariant up to an additive constant
\be
\tilde\S_0(\tilde q)=\S_0(q).
\l{abbgth}\ee
Note that
\be
\tilde\T_0(\tilde p)=\T_0(p)+\rho^{-1}{(AC} pq^2+{BD} p+2BC pq).
\l{hfgy}\ee
The transformations (\ref{4}) are equivalent to ($\epsilon=\pm \sqrt{1/\rho}$)
\be
\tilde q\sqrt{\tilde p} =\epsilon(A q\sqrt p +B \sqrt p),
\quad \sqrt{\tilde p}=\epsilon(C q\sqrt p +D \sqrt p),
\l{ssss}\ee
and can be seen as a rotation of the elements in the kernel of a second--order
operator. The second derivative of (\ref{lt}) with respect to $\s=\S_0(q)$ gives
the ``canonical equation''
\be
\left(\partial^2_\s+\U(\s)\right)q\sqrt p
=0=\left(\partial^2_\s+\U(\s)\right)\sqrt p,
\l{10}\ee
where ${\U}(\s)=\{q,\s\}/2$, with $\{h(x),x\}={{h{'''}}/{h{'}}}-
{(3/2)}({h{''}/{h{'}}})^2$ denoting the Schwarzian derivative.

The above method, used in the framework of the Schr\"odinger equation in
\cite{FM}, has been introduced in \cite{m1} for deriving the inversion formula
in $N=2$ super Yang--Mills, and has been further investigated in \cite{Carroll}.

Involutivity of the Legendre transformation and the duality
\be
\S_0\longleftrightarrow\T_0,\quad ~~~{q}\longleftrightarrow p,
\l{stpq}\ee
imply another $GL(2,{\CC})$--symmetry, with the dual versions of (\ref{10})
being ($\t=\T_0(p)$)
\be
\left(\partial^2_\t+\V(\t)\right)p\sqrt q=0=\left(\partial^2_\t+\V(\t)
\right)\sqrt q,
\l{abbdual}\ee
where $\V(\t)=\{p,\t\}/2$. Now observe that in the case in which $p=\gamma/q$,
the solutions of (\ref{10}) and (\ref{abbdual}) coincide and $\U(\s)=-{1/4
\gamma^2}=\V(\t)$. We will call self--dual the states parametrized by $\gamma$.
These states correspond to
\be
\S_0=\gamma\ln \gamma_q q,\qquad\T_0=\gamma\ln\gamma_p p.
\l{selfduals}\ee
Since
\be
\S_0+\T_0=pq=\gamma,
\l{s0t0}\ee
it follows that the dimensional constants $\gamma_p$, $\gamma_q$ and $\gamma$,
satisfy the relation
\be
\gamma_p\gamma_q\gamma=e.
\l{gammasss}\ee

We observe that as $q$ and $p$ above are not considered independent, the
transformations in (\ref{4}) are not canonical ones. Nevertheless, note that,
as in the search for canonical transformations leading to a system with
vanishing Hamiltonian one obtains the Hamilton--Jacobi equation, we may
similarly look for the equation one obtains by considering the transformation
of $q$, which induces the transformation of the dependent variable $p=\partial_q
\S_0(q)$, reducing to the free system with vanishing energy. Answering
this basic question will lead to the formulation of the equivalence principle
and then to the quantum analogue of the Hamilton--Jacobi equation.

Let us first generalize (\ref{4}) to arbitrary coordinate transformations
$q\to \tilde q(q)$. Note that setting $\tilde \S_0(\tilde q)=\S_0(q(\tilde q))$
is a natural way to associate a new reduced action to the coordinate
transformation. As $\tilde p=\partial_{\tilde q}\tilde\S_0(\tilde q)$, it
follows that in passing from $\S_0$ to $\tilde\S_0$, $p$ transforms as
$\partial_q$. Similarly, the dual version $q\sim \partial_p$ arises by
associating to an arbitrary transformation $p\to \tilde p(p)$ the state $\tilde
\T_0$ defined by $\tilde\T_0(\tilde p)=\T_0(p(\tilde p))$.

Under $q\to \tilde q(q)$, the associated Legendre transformation $\tilde \S_0
(\tilde q)=\tilde p\partial_{\tilde p}\tilde\T_0(\tilde p)-\tilde\T_0(\tilde p)$
generates Eq.(\ref{10}) with the ``canonical potential'' $\tilde \U(\tilde{\s}
)$. While for a M\"obius transformation both $\tilde q\sqrt{\tilde p}$ and
$\sqrt{\tilde p}$ are by (\ref{ssss}) still solutions of (\ref{10}), so that
$\tilde \U(\tilde\s)=\U(\s)$, this is no longer the case for arbitrary
coordinate transformations. This is a consequence of the properties of the
Schwarzian derivative, as $\{\tilde q,\s\}=\{q,\s\}$ if and only if
$\tilde q=(Aq+B)/(Cq+D)$ \cite{Nehari}. Observe that for a given $\U$, the ratio
of two linearly independent solutions of (\ref{10}) gives, up to a M\"obius
transformation, $q=f(\s)$. Inverting it we get the solution of the equation
of motion $\s=\S_0(q)$. Hence, states with the same $\U$ correspond to
specifying different initial conditions of (\ref{10}). However, under arbitrary
transformations we have $\tilde\U(\tilde\s)\ne\U(\s)$, unless one considers the
transformations (\ref{4}). It follows that different $\U$'s can be connected by
coordinate transformations. Similarly, as we noticed, two systems $\tilde\S_0$
and $\S_0$ are related by the transformation $q\to\tilde q(q)$ defined by
$\tilde\S_0(\tilde q)=\S_0(q(\tilde q))$. Therefore, to find the coordinate
transformation connecting $\S_0$ with $\tilde\S_0$ is equivalent to solving the
inversion problem
\be
q\longrightarrow\tilde q=\tilde \S_0^{-1}\circ \S_0(q).
\l{thebasicidea}\ee
This suggests the following ``equivalence principle":

\vspace{.333cm}

\noindent
{\it For each pair $\W^a,\W^b$ there is a coordinate transformation such that
$\W^a(q)\to \tilde \W^a(\tilde q)=\W^b(\tilde q)$.}

\vspace{.333cm}

Observe that this implies that there always exists a coordinate transformation
reducing to $\W=0$ corresponding to the free system with vanishing energy.

We now show the basic fact that this principle is not consistent with classical
mechanics. Let us consider the CSHJE
\be
{1\over 2m}\left({\partial\S_0^{cl}(q)\over \partial q}\right)^2+V(q)-E=0.
\l{clls}\ee
Under (\ref{4}) we have ${\partial_{\tilde q} \tilde\S_0^{cl}(\tilde q)}=
(Cq+D)^2{\partial_{q}\S_0^{cl}(q)}$ and $\tilde\U(\tilde\s^{cl})=\{\tilde
q,\tilde\s^{cl}\}/2=\U(\s^{cl})$. On the other hand, as ${1\over 2m}(
{\partial_{\tilde q} \tilde\S_0^{cl}(\tilde q)})^2+\tilde \W(\tilde q)=0$,
consistency, {\it i.e.} covariance, implies that $\tilde \W(\tilde q)=(Cq+D)^4
\W(q)$. Similarly, in the case of $v^{cl}$--transformations $q\to \tilde q=
v^{cl}(q)$, defined by $\tilde\S_0^{cl}(\tilde q)=\S_0^{cl}(q(\tilde q))$,
the state corresponding to $\tilde \W$ associated to $\tilde\S_0^{cl}$,
satisfies $\tilde {\W}(\tilde q)(d\tilde q)^2={\W}(q)(dq)^2$. In other words
${\W}(q)$ would belong to $\Q^{cl}$, the space of functions transforming as
quadratic differentials under $v^{cl}$--transformations. It follows that
\be
\W(q)=0\to\tilde\W(\tilde q)=\left(\partial_q \tilde q\right)^{-2}\W(q)=0,
\l{qddd}\ee
that is, due to the homogeneity of the transformation properties of the
quadratic differentials, the state corresponding to $\W=0$ is a fixed point in
the space $\H$ of all possible $\W$. In other words, in classical mechanics the
space $\H$ cannot be reduced to a point upon factorizing by the
$v^{cl}$--transformations.

In the following we will derive a differential equation for $\S_0$
with the following properties
\begin{itemize}\item[{\bf 1.}]{Covariance, {\it i.e.} consistency, under the
$v$--transformations $q\to\tilde q=v(q)$, defined by $\tilde\S_0(\tilde q)=
\S_0(q(\tilde q))$.}
\item[{\bf 2.}]{In a suitable limit it reduces to the CSHJE.}
\item[{\bf 3.}]{All the states $\W\in\H$ are equivalent under the
$v$--transformations.}
\end{itemize}
While point {\bf 1.} is nothing else but a consistency condition and
{\bf 2.} is a consequence of the existence of classical mechanics, point
{\bf 3.} has a highly nontrivial dynamical content as will play the basic role
in fixing the differential equation for $\S_0$.

Without loss of generality, we can write the equation we are looking for in the
form
\be
{1\over 2m}\left({\partial\S_0\over \partial q}\right)^2+\W(q)+Q(q)=0.
\l{aa10bbbb}\ee
Observe that if $\Q$ denotes the space of functions transforming as quadratic
differentials under the $v$--transformations, then as ${1\over 2m}\left(
\partial_{\tilde q}\tilde\S_0(\tilde q)\right)^2+\tilde\W(\tilde q)+\tilde Q(
\tilde q)=0$ we have by consistency that $(\W+Q)\in \Q$. On the other hand,
Eq.(\ref{qddd}) and point {\bf 3.} imply that $\W\notin\Q$, so that we also have
$Q\notin\Q$. We also note that the classical limit $Q\to 0$, for which
Eq.(\ref{aa10bbbb}) reduces to Eq.(\ref{clls}), corresponds to the covariance
breaking limit, so that $Q$ has the geometrical nature of a covariantizing term.

Let us now consider the free system with vanishing energy. In this case
Eq.(\ref{aa10bbbb}) becomes $\left(\partial_q \S_0\right)^2=-2mQ$. Observe that
as $\left(\partial_q \S_0\right)^2\in \Q$, and $Q\notin \Q$, covariance would
apparently imply $Q=0$ so that $\S_0=cnst$. Therefore, as $\tilde\S_0(\tilde q)
=\S_0(q)$, any choice of coordinates would always give $\tilde \S_0=cnst$, so
that, in contradiction with {\bf 3.}, $\S_0=cnst$ would be a fixed point in the
space $\K$ of all possible $\S_0$. This aspect is related to the existence of
the Legendre transformation. In particular, $\S_0$--$\T_0$ duality holds unless
$\S_0=cnst$ or $\S_0\propto q$, for which the formalism breaks down. On the
other hand, one expects that the basic properties of the equations underlying
physical systems should be independent from the specific system one considers.
In particular, we have that the formalism breaks down for the system
corresponding to $\W=0$. Similarly, whereas $\S_0$--$\T_0$ duality holds for an 
accelerated particle, this would not be the case in its rest frame. We will see
that there is a remarkable mechanism, direct consequence of the equivalence
principle, which solves the above problems.

Let us first introduce the basic identity
\be
\left({\partial\S_0\over \partial q}\right)^2=
{\beta^2\over 2}(\{e^{{2i\over\beta}\S_0},q\}-\{\S_0,q\}),
\l{expoid}\ee
which forces us to use the dimensional constant $\beta$. By (\ref{aa10bbbb})
and (\ref{expoid}) we have
\be
\W(q)={\beta^2\over 4m}(\{\S_0,q\}-\{e^{{2i\over\beta}\S_0},q\})-Q(q).
\l{g1}\ee
Since there is no universal constant in the CSHJE with the dimension of an
action, we see that $\beta$ is the only natural parameter we can use in order
to reach the covariance breaking phase in which $Q=0$.

We now consider the natural solution
\be
Q(q)={\beta^2\over 4m}\{\S_0,q\},
\l{djh}\ee
which we will show in \cite{FM3} to be unique. It follows from (\ref{g1}) and
(\ref{djh}) that
\be
\W(q)= -{\beta^2\over 4m}\{e^{{2i\over\beta}\S_0},q\},
\l{gaa1}\ee
which is equivalent to the differential equation
\be
{1\over 2m}\left({\partial\S_0\over \partial q}\right)^2
+V(q)-E+{\beta^2\over 4m}\{\S_0,q\}=0,
\l{aa10bbbxxxb}\ee
that in the $\beta\to 0$ limit reduces to the CSHJE (\ref{clls}).

Eq.(\ref{gaa1}) and the identities
\be
\partial_x {h'}^{1/2}{h'}^{-1/2}=0=\partial_x {h'}^{-1}{\partial_x}{h'}^{1/2}
{h'}^{-1/2}h,
\l{id34}\ee
and ${h'}^{1/2}{\partial_x}{h'}^{-1}{\partial_x}{h'}^{1/2}={\partial^2_x}+
\{h,x\}/2$, imply
\be
e^{{2i\over\beta}\S_0}={A\psi^D +B\psi\over C\psi^D+D\psi},
\l{dfgtp}\ee
$AD-BC\ne 0$, where $\psi^D$ and $\psi$ are linearly independent solutions
of the stationary Schr\"odinger equation
\be
\left[-{\beta^2\over 2m}{\partial^2\over \partial q^2}+V(q)\right]\psi=E\psi.
\l{yz1xxxx4}\ee
Thus, for the ``covariantizing parameter'' $\beta$ we have
\be
\beta=\hbar,
\l{PlanckErwin}\ee
where $\hbar=h/2\pi$ with $h$ the Planck constant.

The formulation manifests explicit $\S_0$--$\T_0$ duality as both $\S_0=cnst$
and $\S_0\propto q$ do not belong to $\K$. Due to the M\"obius invariance of the
Schwarzian derivative \cite{Nehari}, instead of $\S_0=\sqrt{2mE}q$,
corresponding to $\W(q)=-{\hbar^2\over 4m}\{e^{{2i\over\hbar}\S_0},q\}=-E\ne 0$,
we can choose
\be
\S_0={\hbar\over 2i}\ln {Ae^{{2i\over\hbar}\sqrt{2mE}q}+B\over
Ce^{{2i\over\hbar}\sqrt{2mE}q}+D},
\l{alluce2}\ee
where the constants are chosen in such a way that $\S_0\not\propto q$.

For $\W=0$, the equation $\left({\partial_q\S_0}\right)^2=-{\hbar^2}\{\S_0,q\}
/2$, (by (\ref{expoid}) equivalent to $\{e^{{2i\over\hbar}\S_0},q\}=0$) has the
solutions $\S_0={\hbar\over 2i}\ln(Aq+B)/(Cq+D)$. We therefore have the
important fact that $\S_0$ is never a constant! Comparing with (\ref{selfduals})
and relaxing the reality condition on $\S_0$, we can choose for $\W=0$ the pair
of self--dual states
\be
\S_0^{sd}=\pm{\hbar\over 2i}\ln\gamma_q q,
\l{alluce1}\ee
that for $\hbar\to 0$ reduce to the classical result. Physical solutions for
$\S_0$ correspond to values of $A,B,C,D$ in (\ref{dfgtp}) such that $\S_0$ is
real, that is we have
\be
e^{{2i\over\hbar}\S_0}=e^{i\alpha}{\psi^D+i\bar\ell\psi\over\psi^D-i\ell\psi},
\l{iodhO}\ee
where $\alpha\in{\RR}$ and ${\rm Re}\,\ell\ne 0$. Thus, while (\ref{alluce1})
is a complex solution and corresponds to the state with $\W=0$, the physical
solution, still corresponding to $\W=0$, is given by
\be
e^{{2i\over\hbar}\S_0}=e^{i\alpha}{q+i\bar\ell\over q-i\ell}.
\l{iodhO2}\ee

The above analysis shows that while in the standard approach the solutions
corresponding to the state with $\W=cnst$ coincide with the classical ones, here
we have a basic difference related to the existence of the Legendre
transformation of $\S_0$ for any system. The solutions
(\ref{alluce2})(\ref{alluce1}) have been overlooked in the literature.
Note that by (\ref{dfgtp}) the general solution of (\ref{yz1xxxx4}) is
\be
\psi={1\over\sqrt{\S_0'}}\left(A e^{-{i\over\hbar}\S_0}+Be^{{i\over\hbar}\S_0}
\right),
\l{popca}\ee
that for (\ref{alluce2})(\ref{alluce1}) gives, as it should, the solutions
$\psi=Ae^{-{i\over \hbar}\sqrt{2mE}q}+Be^{{i\over \hbar}\sqrt{2mE}q}$ and
$\psi=Aq+B$.

Let us compare the above equations with those of the standard notation
\cite{LL}. While Eq.(\ref{aa10bbbxxxb}) is written in terms of $\S_0$ only,
substituting $\psi=R{\rm exp}(i{\hat\S_0}/\hbar )$ in (\ref{yz1xxxx4}) yields
\be
({\partial_q {\hat\S_0}})^2/2m+V(q)-E-{\hbar^2}{(\partial^2_q R)/2m R}=0,
\l{yz11}\ee
\be
{\partial_q}( R^2 \partial_q{\hat\S}_0)=0.
\l{yz12}\ee
We can distinguish the cases $\overline\psi\not\propto \psi$ and $\overline
\psi\propto \psi$. In the first one we can choose $\psi^D=\overline\psi$, $i.e.$
\be
\psi(q)=R(q)e^{{i\over\hbar}\hat\S_0(q)},\qquad \psi^D(q)=R(q)
e^{-{i\over\hbar}\hat\S_0(q)},
\l{e10}\ee
so that we can set $\S_0=\hat\S_0$. The continuity equation (\ref{yz12}) gives
$R={1/\sqrt{\S_0'}}$ so that $Q(q)=\hbar^2\{\S_0,q\}/4m=-\hbar^2 (\partial^2_q
R)/2m R$ and Eq.(\ref{yz11}) corresponds to Eq.(\ref{aa10bbbxxxb}).

In the $\overline\psi\propto\psi$ case one has that $\hat\S_0$ is a constant,
and we can set $\hat\S_0=0$. This fact shows that identifying the wave function
with $R{\rm exp}(i{\hat\S_0}/\hbar )$, typical of Bohmian mechanics, is
problematic as it would imply a rather involved classical limit. Since the case
$\overline\psi\propto\psi$ corresponds to bound states, we would have systems,
such as the harmonic oscillator, in which in the $\hbar\to 0$ limit one has to
recover a nontrivial classical reduced action from $\hat\S_0=0$. This fact can
be seen as further evidence that the quantum analogue of the classical reduced
action is $\S_0$ rather than $\hat\S_0$. This also implies that $Q$ is the
genuine quantum potential rather than $-\hbar^2 (\partial^2_q R)/2m R$.

Let us further consider the $\overline\psi\propto\psi$ case. Since $\hat\S_0=0$,
we have
\be
\psi(q)=R(q).
\l{oihkl}\ee
Furthermore, Eqs.(\ref{yz11}) and (\ref{yz12}) give
\be
V(q)-E-{\hbar^2}{(\partial^2_q R)/2m R}=0,
\l{iodx}\ee
so that in this case the relation between the standard quantum potential
\be
\hat Q=-{\hbar^2\over 2m} {\partial^2_q R\over R},
\l{oldQP}\ee
and $Q$ is
\be
\hat Q=Q+{1\over 2m}\left({\partial\S_0\over\partial q}\right)^2.
\l{pojdx}\ee

The existence of the self--dual state makes it possible to find a coordinate
$\tilde q$, solution of the Schwarzian equation
\be
\{\tilde q,q\}+{4m}(V(q)-E)/\hbar^2=0,
\l{nonlinearschw}\ee
in which any $\W\in \H$ reduces to $\tilde\W(\tilde q)=0$. In complete analogy
with the fact that the existence of the classical trivializing conjugate
variables $(Q,P)$, defined by the canonical transformation
\be
q \longrightarrow Q,\qquad\qquad p\longrightarrow P=cnst=
-\partial_Q\S_0^{cl}(q,Q)|_{Q=cnst},
\l{canonicalusual}\ee
implies the CSHJE $H(q,\partial_q\S_0^{cl})-E=\tilde H(Q,P)=0$, we have that
(\ref{nonlinearschw}) is a consequence of the existence of the trivializing map
\be
q\longrightarrow \tilde q= {\gamma_q}^{-1}e^{{2i\over\hbar}\S_0},\qquad\qquad
p\longrightarrow  \tilde p=(\partial_q \tilde q)^{-1}p={i\hbar/ 2\tilde q},
\l{duyqg}\ee
leading to the free system with vanishing energy. Eq.(\ref{duyqg}) is the
solution of the inversion problem (\ref{thebasicidea}) when $\tilde \S_0$ is
the reduced action of the state with $\tilde\W=0$. Therefore, given an
arbitrary state $\W\in\H$, the transformation (\ref{duyqg}) gives $\tilde
\W(\tilde q)=0$, and by (\ref{popca}) Eq.(\ref{yz1xxxx4}) becomes $(\partial_q
\tilde q)^{3/2}\partial_{\tilde q}^2\tilde\psi(\tilde q)=0$, where
\be
\tilde \psi (\tilde q)(d\tilde q)^{-1/2}=\psi (q)(dq)^{-1/2}.
\l{23}\ee

We note that the trivializing map can be transformed to a real map by performing
a Cayley transformation of $e^{{2i\over\hbar}\S_0}$. Since this map is a
M\"obius transformation, it is a symmetry of $\W=-\hbar^2\{e^{{2i\over\hbar}
\S_0},q\}/4m$.

Remarkably, the quantum correction to the CSHJE (\ref{clls}), can be also seen
as modification by a ``conformal factor'' defined by the canonical potential
\be
{1\over 2m}\left({\partial\S_0\over\partial q}\right)^2
\left[1-\hbar^2 \U(\S_0)\right]+V(q)-E=0,
\l{hsxdgyij}\ee
where we used the identity $\{q,\S_0\}=-(\partial_q\S_0)^{-2}\{\S_0,q\}$.
This shows the basic role of the purely quantum mechanical self--dual states
(\ref{alluce1}) as in this case
\be
1-\hbar^2\U(\pm {\hbar\over 2i}\ln \gamma_qq)=0,
\l{conformee}\ee
which are two possible solutions of (\ref{hsxdgyij}) for $\W=0$, the other
possible solutions are given by $S_0={\hbar\over 2i}\ln (Aq+B)/(Cq+D)$,
$AD-BC\ne 0$.

We note that an additional term in (\ref{djh}) would imply a differential
equation for $\S_0$ which could not satisfy conditions {\bf 1.--3.} \cite{FM3}.

We observe that by (\ref{23}) it follows that, in general, diffeomorphisms do
not preserve the transition amplitudes and are not unitary. This is of course
expected as these transformations connect any pair of different physical
systems.

We have seen that the requirement of preserving the original structures observed
in the Legendre transformation can be consistently satisfied. This brings us to
the Schr\"odinger equation which can be characterized by the equation $\W(q)=-{
\hbar^2}\{e^{{2i\over\hbar}\S_0},q\}/4m$, or equivalently by the term $Q(q)={
\hbar^2}\{\S_0,q\}/4m$ which added to the CSHJE leads to the Schr\"odinger
equation. Recalling the structure of the canonical potential, namely $\U(\S_0)=
\{q,\S_0\}/2$, we explicitly see how the basic M\"obius symmetry, a
characteristic property of the Schwarzian derivative, still survives in quantum
theory. Thus, the canonical formalism by itself unavoidably contains an
intrinsic M\"obius ambiguity which actually turns out to be at the heart of
quantum mechanics. In particular, the fact that the relevant equations remain
invariant under M\"obius transformations of the canonical variables and the
related existence of the self--dual states, characterized by $\gamma=\pm\hbar
/2i$, reflect in the reconsideration of these classical variables.

We stress that an important aspect in our construction concerns the identity
(\ref{expoid}) which contains both the classical and quantum parts $\W$ and $Q$
respectively. In particular, note that it includes in the same equation both
$e^{{2i\over\hbar}\S_0}$ and $\S_0$. If one considers $\S_0$ as a scalar field
operator, then the ``vertex'' $e^{{2i\over\hbar}\S_0}$ resembles the
bosonization of a fermionic operator. It is amusing that inspired by duality
in SUSY Yang--Mills \cite{FM,m1}, we obtained a quantum mechanical expression
reminiscent of supersymmetry.

Though it may seem specifically one--dimensional, our formulation implies
quantum mechanics also in higher dimensions \cite{BFM}. This is just like the
Heisenberg uncertainty relations $\Delta p_k\Delta q_k\geq \hbar/2$, which, in
spite of being intrinsically one--dimensional, actually encode quantum mechanics
in any dimension. In particular, since the formulation trivially extends to the
case when
\be
V(q)=\sum_{k=1}^DV_k(q_k),
\l{separa}\ee
we have that the state with $\W=0$ still corresponds to the nontrivial universal
solution
\be
\S_0^{sd}=\pm{\hbar\over 2i}\sum_{k=1}^D\ln \gamma_qq_k.
\l{nonttt}\ee
This guarantees that the Legendre transformation
\be
\S_0=\sum_{k=1}^Dp_k{\partial \T_0\over \partial p_k}-\T_0,
\l{podl}\ee
is defined for any physical system and, as in the one--dimensional case, its
involutivity implies $\S_0$--$\T_0$ duality. Therefore, in higher dimensions one
should derive an equation that, for potentials of the form (\ref{separa}), is
equivalent to decoupled one--dimensional Schr\"odinger equations. Furthermore,
in the classical limit it should reproduce the CSHJE. In \cite{BFM} it will be
shown how these conditions yield the Schr\"odinger equation in any dimension.

Finally, in the time--dependent case the equation for the action $\S$ is
determined by considering that in the classical limit it should correspond to
the Hamilton--Jacobi equation and that in the time--independent case it
reproduces the above results. This implies the quantum Hamilton--Jacobi equation
in the general case and then the time--dependent Schr\"odinger equation
\cite{BFM}.

\vspace{1cm}

\noindent
{\bf Acknowledgments}. We would like to thank G. Bertoldi, G. Bonelli, L.
Bonora, R. Carroll, G.F. Dell'Antonio, E.S. Fradkin, E. Gozzi, F. Illuminati,
A. Kholodenko, A. Kitaev, P.A. Marchetti, J. Ng, P. Pasti, S. Shatashvili, M.
Tonin and R. Zucchini, for discussions. Work supported in part by DOE Grant
No.\ DE--FG--0586ER40272 (AEF) and by the European Commission TMR programme
ERBFMRX--CT96--0045 (MM).

\end{document}